\documentclass[5p,twocolumn,sort&compress]{elsarticle}

\usepackage{lineno}
\modulolinenumbers[5]
\usepackage{amssymb}
\journal{Journal of Magnetic Resonance}

\usepackage[export]{adjustbox} % orients graphics left, right.
\usepackage{graphicx }% include figure files
\usepackage{textgreek}% greek letters in text
\usepackage[]{hyperref}
\hypersetup{colorlinks=true,linkcolor=blue,citecolor=blue,urlcolor=blue,pdfpagemode=UseNone}

\newcommand{\slrr}      {$T_1^{-1}$}

\begin{document}

%\title{Evaluation of Inverse Laplace Transformation Analysis of NMR Magnetization Data}
\title{Inverse Laplace Transformation Analysis of Stretched Exponential Relaxation}

\author{H. Choi\fnref{myfootnote}}
\address{College of Nanoscale Science and Engineering, State University of New York Polytechnic Institute, New York 12203, USA}
\fntext[myfootnote]{Department of Materials Science and Engineering, Northwestern University, Evanston, IL 60208, USA;
hchoi@u.northwestern.edu}

\author{I. Vinograd}
%\address{Department of Physics, University of California, Davis, California
%95616, USA}

\author{C. Chaffey}
%\address{Department of Physics, University of California, Davis, California
%95616, USA}

\author{N. J. Curro}
\address{Department of Physics, University of California, Davis, California
95616, USA}
\date{\today}

\begin{abstract}
We investigate the effectiveness of the Inverse Laplace Transform (ILT) analysis method to extract the distribution of relaxation rates from nuclear magnetic resonance data with stretched exponential relaxation.  Stretched-relaxation is a hallmark of a distribution of relaxation rates, and an analytical expression exists for this distribution for the case of a spin-1/2 nucleus. We compare this theoretical distribution with those extracted via the ILT method for several values of the stretching exponent and at different levels of experimental noise.  The ILT accurately captures the distributions for $\beta \lesssim 0.7$, and for signal to noise ratios greater than $\sim 40$; however the ILT distributions tend to introduce artificial oscillatory components.  We further use the ILT approach to analyze stretched relaxation for spin $I>1/2$ and find that the distributions are accurately captured by the theoretical expression for $I=1/2$. Our results provide a solid foundation to interpret distributions of relaxation rates for general spin $I$ in terms of stretched exponential fits.
%in providing estimations of the probability distribution of spin-lattice relaxation rates P(W\textsubscript{1}) from nuclear magnetic resonance (NMR) magnetization recovery data. To evaluate this method, P(W\textsubscript{1}) obtained using the ILT analysis method are compared to corresponding analytic solutions, which are defined for recovery functions with a stretched exponential form. For further study of the method, parameters such as the number of points and noise for the recovery curves are varied to observe their effect on the estimated P(W\textsubscript{1}). We observe that these parameters contribute to undesired oscillatory behavior in the estimations.
\end{abstract}


\maketitle

%\pacs{75.30.Gw,75.40.Cx,71.20.−b, 76.60.-k}% PACS, the Physics and Astronomy
\section{Introduction}

One of the most important quantities measured in magnetic resonance is the spin lattice relaxation rate, \slrr.  This quantity probes the interaction between the nuclear spins and their environment, and reveals information about the local dynamics at the nuclear spin site. In conductors, the dominant contribution to \slrr\ arises from the hyperfine coupling to the electron spins, and in this case \slrr\ can be directly related to the dynamical spin susceptibility of the electrons \cite{MoriyaT1formula}.  This relationship has been extensively utilized to study a number of correlated electron systems, ranging from high temperature superconducting cuprates \cite{TakigawaONMRinYBCO,Y1248T1SCstate,ImaiLSCO,Ohsugi,Julien1996} , heavy fermion materials \cite{Urbano2007,UPt3T1study,ZhengCeRhIrIn5}, iron-based superconductors \cite{takigawa2008,TakigawaSr122pressure,Kissikov2018} and other exotic materials \cite{UPt3T1study,Baek2010a}.  A common issue in correlated electron systems is the presence of electronic inhomogeneity \cite{Tranquada1995,Kivelson1998,AlloulHirschfeld,schmalianglass,ParkDropletsNature2013,DioguardiNematicGlass2015}.  Even in a single crystal, inhomogeneity may arise intrinsically via frustration among competing orders, from the presence of impurities, or via inhomogeneous electronic responses such as in the mixed state of a type II superconductor.  If the inhomogeneity is purely static, then the nature of the inhomogeneous distribution can often be studied via the effect on the NMR spectrum \cite{HorvaticCuGeO3NMR,Nakai2008a}. If the inhomogeneity fluctuates, then the spectrum may be motionally narrowed, precluding such investigations.  However, in many cases the dynamics of the inhomogeneity may be reflected in \slrr, in which case there is a \emph{distribution of relaxation rates},  rather than a single homogeneous \slrr.

If each nucleus in a crystal relaxes with a different relaxation rate, $W_1(\mathbf{r})$, where $\mathbf{r}$ describes the position of the nucleus in the lattice,  then the total magnetization measured experimentally exhibits a complicated relaxation curve.  If the distribution of relaxation rates, $P(W_1)$, is sufficiently narrow, then the NMR magnetization recovery data can be fit to a exponential form, $\exp[-t/T_1]$ (for a spin $I=1/2$ nucleus), where $t$ is the recovery time, and \slrr\ is the median of $P(W_1)$. On the other hand, if $P(W_1)$ is wide, then this exponential form will not fit the magnetization recovery data well, making it difficult to extract a meaningful value of \slrr.  It is common practice to fit the recovery to a stretched exponential form, $\exp[-(t/T_1)^{\beta}]$, where $\beta$ is the so-called stretching exponent that provides a rough measures of the width of $P(W_1)$, and satisfies $0<\beta \leq 1 $  \cite{Phillips1996,johnstonstretched}. In many glassy systems, it has been observed that $\beta$ decreases from unity as the temperature is reduced \cite{MitrovicGlassy214PRB2008,DioguardiPdoped2015}.  As $\beta$ is reduced, the distribution $P(W_1)$ grows by several decades, and $\beta$ is related to the logarithmic width of the distribution. Analyzing magnetization recovery data with a stretched exponential is straightforward to implement and requires only one extra fitting parameter.

A significant disadvantage to this approach, however, is that it makes an implicit assumption about the shape of $P(W_1)$ that may or may not accurately reflect the true distribution.  A more direct method to extract the distribution is desirable, but is in fact  an ill-posed problem and is non-trivial to implement.  The magnetization decay curve,  $M(t)$,  is related to $P(W_1)$ via a Fredholm integral of the first kind \cite{butler1981estimating,honerkamp1990tikhonovs}.  For a spin-1/2 nucleus, $P(W_1)$ reduces to the inverse Laplace transform of $M(t)$.  Such problems are notoriously difficult because even small experimental errors in the values of $M(t)$ give rise to large variations in $P(W_1)$, and there is often no unique solution for a given data set.  Different approaches have been developed to extract $P(W_1)$, such as the maximum entropy method \cite{livesey1987analyzing}, and via linearization methods such as Tikhonov regularization \cite{tikhonov1963regularization}.  The latter approach was adopted early on by researchers studying pore size distribution of rocks in the petrochemical industry \cite{fordham1995imaging,venkataramanan2002solving,Song2002,Mitchell2012}, for the investigation of dielectric spectra in glasses \cite{DielectricRegularization}, and recently has been used to analyze the glassy NMR behavior of high temperature superconductors \cite{arsenault2020magnetic,singer2020LBCO}.  This technique holds promise to shed light on many physical systems of interest, but several questions concerning the limits of validity of this approach remain outstanding.   To better understand these limits, we have conducted numerical studies comparing the inverse Laplace transform (ILT) for stretched exponential decays for several different nuclear spins ($I=1/2, \cdots, 9/2$), different levels of signal to noise ratios, and different numbers of measured time points. We find that the ILT algorithm reproduces the theoretical distribution for a spin 1/2 nucleus for small stretching exponents $\beta \leq 0.8$ when the distribution is not narrowly peaked.  For higher spin nuclei, we find that  $P(W_1)$ for stretched relaxation is independent of $I$ as long as the stretched relaxation curve is properly defined.  These results provide important guidance for setting up experiments with sufficient signal to noise to properly extract the distribution of relaxation rates, and for interpreting the distribution when the relaxation can be described by stretched exponentials.

\section{Methods}

When a spin $I=1/2$ nucleus at lattice position $\mathbf{r}$ is not in thermal equilibrium, the magnetization component along the quantization axis (typically the magnetic field direction) relaxes as:
\begin{equation}
    m_z(\mathbf{r},t) = m_0 \left(1-\phi e^{- W_1(\mathbf{r}) t}\right),
\end{equation}
where $m_0$ is the equilibrium magnetization, and $\phi$ is a parameter that describes the initial condition and depends on the pulse sequence employed in the measurement.  We assume that $W_1(\mathbf{r})$ depends on position, $\mathbf{r}$.  All of the nuclei contribute to the measured signal, so that
\begin{eqnarray}
\nonumber M(t) & = & \int_V m_z(\mathbf{r},t) d\mathbf{r}\\
& = &   M_0 \int_0^{\infty} K (W_1, t) P(W_1) dW_1,
\end{eqnarray}
where $V$ is the volume of the sample, and $M_0 = N_0 m_0$, where $N_0$ is the number of nuclei in the crystal. In the second line, rather than integrating over real space we express the integral as a distribution over a normalized distribution of $W_1$ and kernel function $K(W_1, t) = 1-\phi e^{- W_1 t}$.  This kernel function changes for higher spins, $I>1/2$, as described below, however the general approach to solving for $P(W_1)$ remains the same.  For $I=1/2$,
\begin{equation}
    (M(t) - M_0)/\phi =  \int_{0}^{\infty} e^{-W_1 t}P(W_1)dW_1
\label{eqn:magrec}
\end{equation}
is equivalent to the Laplace transform of $P(W_1)$.  Thus in principle, the distribution can be obtained by simply taking the \emph{inverse Laplace transform} of the measured data. The ILT approach offers a powerful method to determine $P(W_1)$, however it requires a number of assumptions.  To determine the distribution, the problem is first linearized:
\begin{equation}
    M_i = M(t_i) = \sum_j K_{ij} P_j + e_i
\end{equation}
where $i \in \{1, \cdots,  N\}$ are the measured time points, $K_{ij} = K(W_{1,j}, t_i)$, $e_i$ are experimental errors, $P_j = P(W_{1j})$,  and $j\in \{1, \cdots, L\}$
with $L>N$ are the points in the distribution. Since $L>N$, there are in fact more points in the distribution than experimentally measured, and the vector $\vec{P}$ is underdetermined. \emph{Tikhonov regularization} \cite{tikhonov1963regularization} offers a method to obtain a solution by minimizing the functional:
\begin{equation}
  \Phi(\vec{P}\,) = \frac{1}{2}|\tilde{\mathbf{K}}\cdot\vec{P} - \vec{M}|^2 + \frac{1}{2}\alpha\left|\vec{P}\right|^2
  \label{eqn:functional}
\end{equation}
subject to the condition that every element $P_j \geq 0$.
Here $\alpha$ is the Tikhonov regularization parameter that enforces $\vec{P}$ to have a stable solution. This procedure ensures that the distribution is positive definite, hence physically realistic, but has the effect of broadening and smoothing the distribution, depending on the choice of $\alpha$ \cite{honerkamp1990tikhonovs,lawson1995solving}.  The solution of (\ref{eqn:functional}) is $\vec{P} = \tilde{\mathbf{H}}\cdot\tilde{\mathbf{K}}^{\dagger}\cdot \vec{c} $, where $\dagger$ means transpose, and the matrix $\tilde{\mathbf{H}}$ has elements $H_{ij} = H((\tilde{\mathbf{K}}^{\dagger}\cdot \vec{c}\,)_{j})$ if $i=j$ and 0 otherwise, and  $H(x)$ is the Heaviside function: $H(x) = 1$ if $x>0$ and $H(x) = 0 $ otherwise.  The vector $\vec{c}$ satisfies:
\begin{equation}
    \vec{c}(\alpha) =\left(\tilde{\mathbf{K}}\cdot \tilde{\mathbf{H}}(\vec{c}(\alpha))\cdot\tilde{\mathbf{K}}^{\dagger} +  \alpha\tilde{\mathbf{I}}\right)^{-1}\cdot\vec{M}
\end{equation}
where $\mathbf{I}$ is the identity matrix. This equation can be solved iteratively \cite{singer2020LBCO}. The sum of the residuals is:
\begin{equation}
    \chi(\alpha) = |\vec{M} - \mathbf{K}\cdot \vec{P} | = \alpha |\vec{c}(\alpha)|.
\end{equation}
The distribution clearly depends on the choice of $\alpha$, and becomes broader and smoother as $\alpha$ increases.  The optimal value of $\alpha$ is usually determined by the so-called self-consistency method \cite{singer2020LBCO,honerkamp1990tikhonovs}, in which $\alpha$ is chosen as the minimum of either $\alpha_1$ or $\alpha_2$, where $\alpha_1 = |\vec{e}|$, the sum of the experimental errors of the measurements of $\vec{M}$, and $\alpha_2$ satisfies:
\begin{equation}
    \left.\frac{d \ln\chi(\alpha)}{d\alpha} \right|_{\alpha_2} = 0.1.
\end{equation}
We use the IGOR Pro software environment to solve for $\vec{c}$ numerically using a set of $\vec{M}$ data, for various numbers of $N$ data points, and with $M = 128$ logarithmically-spaced values of $W_1$. By computing $\chi(\alpha)$ for a broad range of $\alpha$, we find the optimal regularization parameter and use this to determine the distribution $\vec{P}$ for a given data set, $\{ t_i, M_i\}$ with measurements errors $e_i$.

\section{Results}

To determine the effectiveness of the ILT method, it is valuable to test the algorithm to extract known distributions from test data sets. It is also instructive to determine the optimal experimental conditions to get the most accurate measurement of $P(W_1)$.  For example, the number of data points in a typical experiment lies between $ N \sim 5 - 20$.  However, the major constraint is the total experimental time, $t_{expt} = N t_1$, where $t_1$ is the measurement time for a single point.  For Gaussian noise, the measurement error $e_i\sim t_1^{-1/2} \sim N^{1/2}$, for a fixed $t_{expt}$, therefore fewer points would result in lower measurement noise.  An interesting question is whether it is better to have more points, $N$, with higher noise, or fewer points with lower noise, in order to determine $P(W_1)$ with the best fidelity.

\subsection{Stretched Relaxation of a Spin 1/2}

\begin{figure}[h]
\begin{center}
\includegraphics[width=\linewidth]{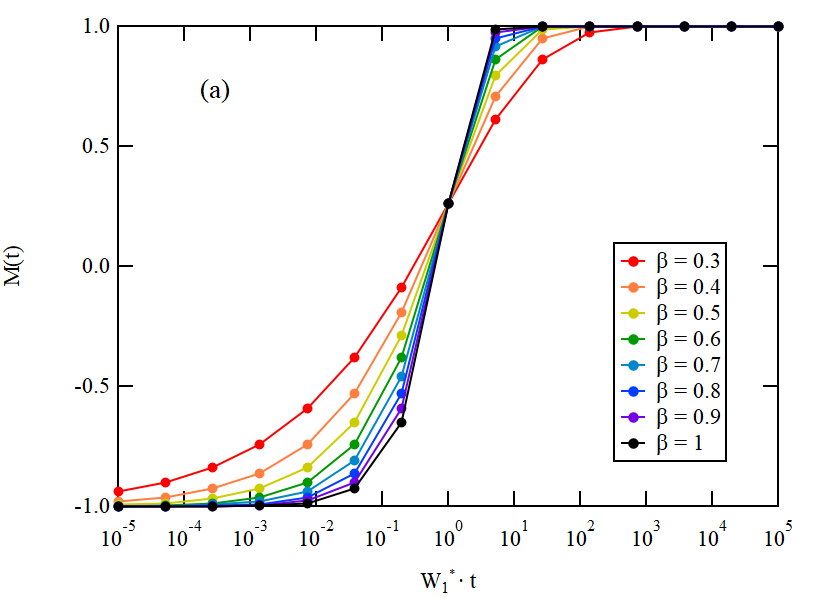}
\includegraphics[width=\linewidth]{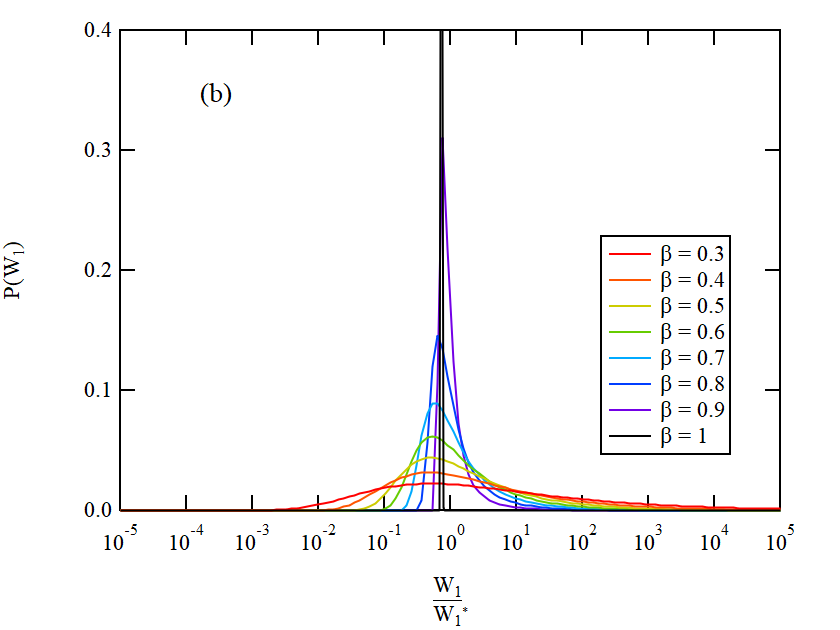}
\caption{\label{fig:stretched} (a) Linear-log plot of spin I = 1/2 M(t) with stretched exponential form given by Eq. \ref{eqn:mtspinhalf} with $N = 15$ time points and $e_i=0$ for $\beta$ ranging from 0.3 to 1 in 0.1 increments. (b) Linear-log plot of the theoretical distribution $P_{\beta}(W_1)$ given by Eq. \ref{eqn:johnstondistribution} with corresponding $\beta$ values.}
\end{center}
\end{figure}

We first consider the case of stretched exponential relaxation of a spin $I=1/2$, with the kernel function:
\begin{equation}
K(t, W_1) = 1 - 2e^{-(W_1t)},
%K(t, W_1) = 1 - 2e^{-(W^*_1t)^\beta}%
\label{eqn:spinhalf}
\end{equation}
with the corresponding magnetization recovery vector:
\begin{equation}
M(t_i) = 1 - 2e^{-(W^*_1t_i)^\beta} + e_i,
\label{eqn:mtspinhalf}
\end{equation}
where $W_1^*$ is a characteristic rate scale. This function is shown in Fig. \ref{fig:stretched}(a) on a linear-log scale with $W_1^* = 1$.
The distribution $P_{\beta}(W_1)$ can be expressed analytically as an infinite series:
\begin{equation}
P_{\beta}(W_{1}) = \frac{1}{\pi}\sum_{n = 1}^{\infty }\frac{(-1)^{n +1}\Gamma (n\beta + 1)}{n!(W_1/W_1^*)^{n\beta + 1}} \sin(n\pi\beta)
\label{eqn:johnstondistribution}
\end{equation}
where $\Gamma(x)$ is the Gamma function \cite{johnstonstretched}. Figure \ref{fig:stretched} shows the distributions corresponding to stretched relaxation for several values of $\beta$. Note that because the $W_1$ values are distributed on a logarithmic scale, it is necessary to multiply $P_{\beta}(W_1)$ by $W_1$ to properly normalize the distribution.  The distributions are centered close to $W_1^*$, which is approximately equal to the median of the distribution. The distribution approaches a delta function as $\beta\rightarrow 1$.  As $\beta$ is reduced, the distribution broadens considerably, and is several decades in width once $\beta \leq 0.5$.

\begin{figure}[h]
\begin{center}
\includegraphics[width=\linewidth]{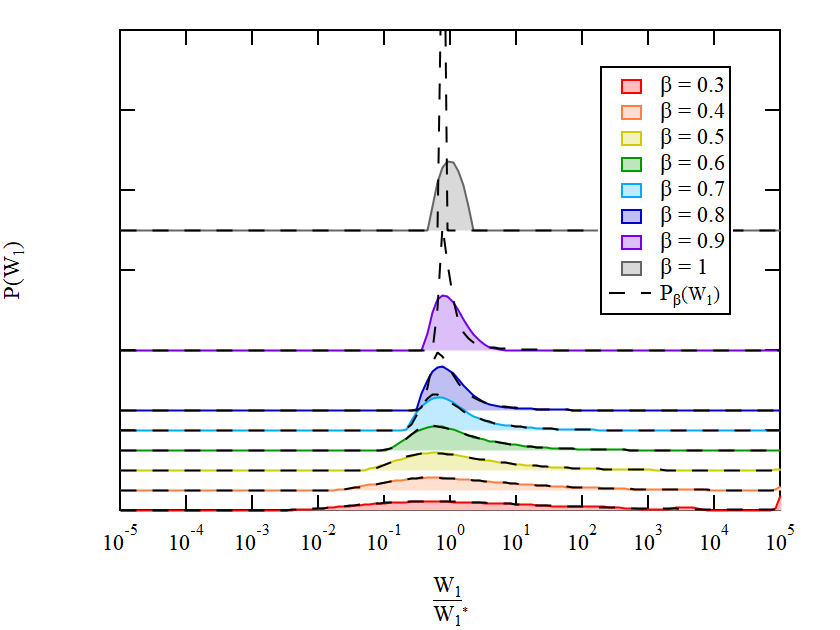}
\caption{\label{fig:comparewaterfall} Linear-log plot of $P(W_1)$ ILT estimations of M(t) given by Eq. \ref{eqn:mtspinhalf} with $N = 15$ time points and $e_i=0$ and the theoretical distribution $P_{\beta}(W_1)$ given by Eq. \ref{eqn:johnstondistribution} for $\beta$ values ranging from 0.3 to 1.0 in 0.1 increments.}
\end{center}
\end{figure}

\begin{figure}[h]
\begin{center}
\includegraphics[width=\linewidth]{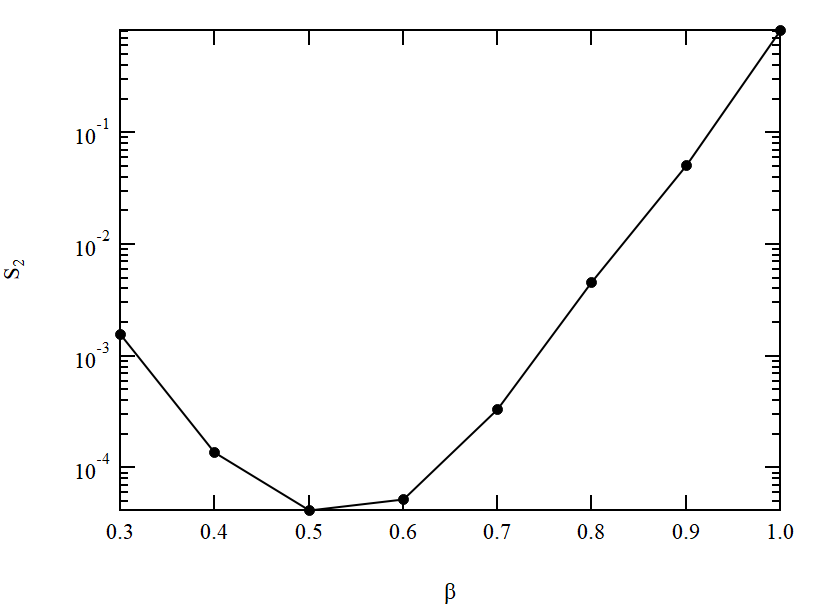}
\caption{\label{fig:comparestretched} Log-linear plot of $S_2$, given by Eq. \ref{S_2}, of P(W\textsubscript{1}) ILT estimations of M(t) given by Eq. \ref{eqn:mtspinhalf} with $N = 15$ time points, and $e_i=0$ for $\beta$ values ranging from 0.3 to 1.0 in 0.1 increments versus $\beta$.}
\end{center}
\end{figure}

Fig. \ref{fig:comparewaterfall} shows the distributions extracted using the ILT method for several values of $\beta$ with zero noise ($e_i=0$) and with $N=15$ time points.  There is relatively good agreement for intermediate values of $\beta$, but once $\beta\gtrsim 0.8$, the distribution narrows and the ILT method fails to capture the narrow width.  To measure the effectiveness of the approach, we compute the sum of the squares of the residuals, $S_2$, defined as:
\begin{equation}
    S_2 = \sum_{j=1}^{M} \left( P_j - P_{\beta}(W_{1j}) \right)^2
\label{S_2}
\end{equation}
where $\vec{P}$ is determined by ILT.  As shown in Fig. \ref{fig:comparestretched}, $S_2$ generally increases as $\beta$ approaches unity. However, $S_2$ has inflated values at $\beta < 0.5$ due to domain constraints. For lower $\beta$ recoveries, the complete magnetization recovery is not captured within the given time domain. As a result, artificial higher relaxation rates are produced in the ILT distributions of lower $\beta$ recoveries, resulting in larger $S_2$ values.

\subsection{Optimal Number of Measurements}
\begin{figure}[h]
\begin{center}
\includegraphics[width=\linewidth]{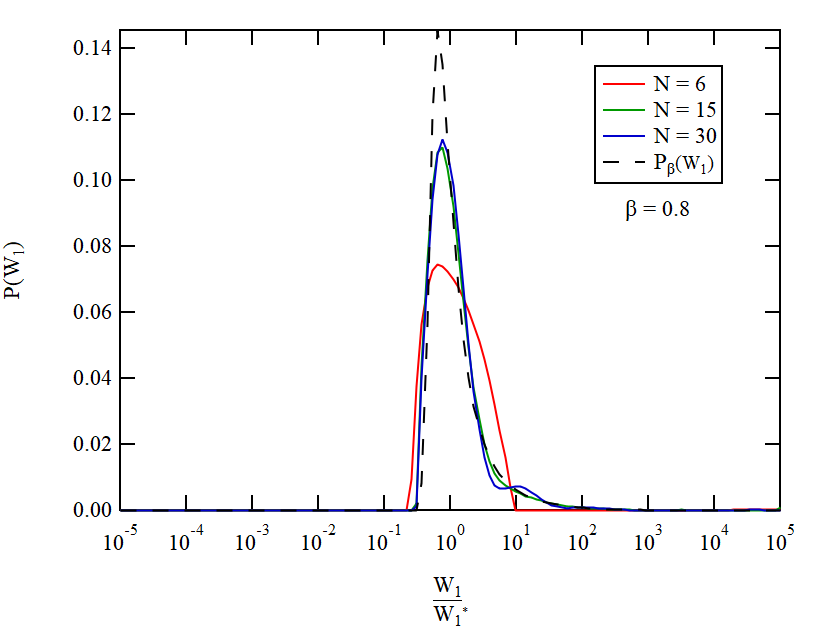}
\caption{\label{fig:distributionvaryN} Linear-log plot of P(W\textsubscript{1}) ILT estimations of M(t) given by Eq. \ref{eqn:mtspinhalf} with $N = 6, 15, 30$ time points and $e_i=0$ and the theoretical distribution $P_{\beta}(W_1)$ given by Eq. \ref{eqn:johnstondistribution} for \textbeta{ } = 0.8.}
\label{fig:T1}
\end{center}
\end{figure}

\begin{figure}[h]
\begin{center}
\includegraphics[width=\linewidth]{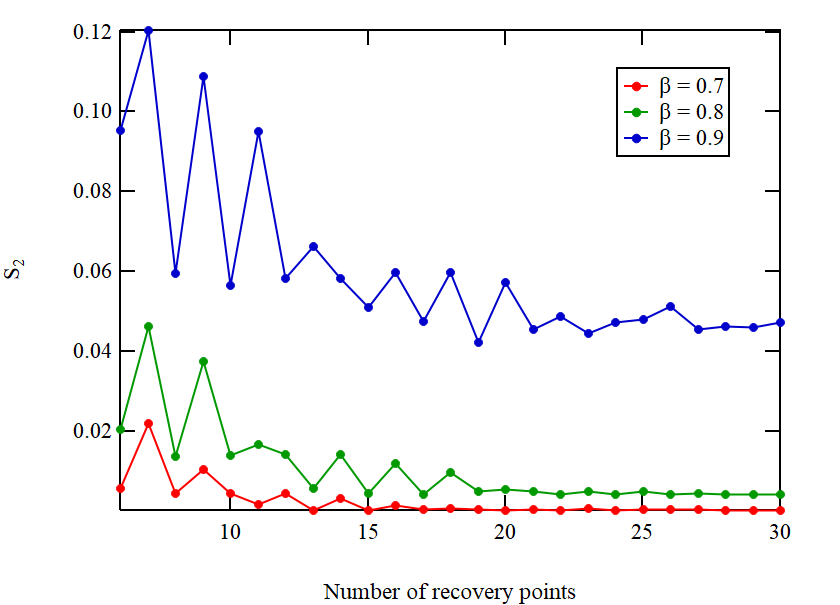}
\caption{Plot of $S_2$, given by Eq. \ref{S_2}, of P(W\textsubscript{1}) ILT estimations of M(t) with $N$ ranging from 6 to 30 time points and $e_i=0$ versus the number of M(t) recovery points for \textbeta{ } = 0.7, 0.8, 0.9.}
\label{fig:varyN}
\end{center}
\end{figure}

To understand how well the algorithm behaves with different numbers of measured time points, we compare the extracted distribution for different values of $N$.  As shown in Fig. \ref{fig:distributionvaryN} for $\beta = 0.8$, including a greater number of measured recovery points improves the quality of the `fit' such that the ILT distribution more accurately reproduces the exact solution.  In each case, the mean of the distribution is correct, but the width is too wide for $N=6$ points.  For 15 points, the agreement is better, but there is an oscillation present in the upper tail of the ILT distribution that is not present in the exact solution.
%Increasing $N$ to 30 points significantly improves this fit to this upper tail, but still fails to capture the magnitude of the sharp peak around $W_1^*$.
The behavior of these oscillations depends on the $N$ when $N\gtrsim 12-15$, but there are no obvious trends. In fact, for $N=30$ the oscillations appear somewhat larger than for $N=15$. To quantify the difference between the ILT and the exact distributions, we compute $S_2$ for various values of $\beta$ and $N$ as shown in Fig. \ref{fig:varyN}. This quantity appears to reach an asymptotic value by approximately $N=12$ to 15.  For smaller values of $N$, $S_2$ oscillates between larger and smaller values for $N$ odd or even values, respectively.  The origin of this behavior is not understood.

\subsection{Sensitivity to Noise}

\begin{figure}[h]
\begin{center}
\includegraphics[width=\linewidth]{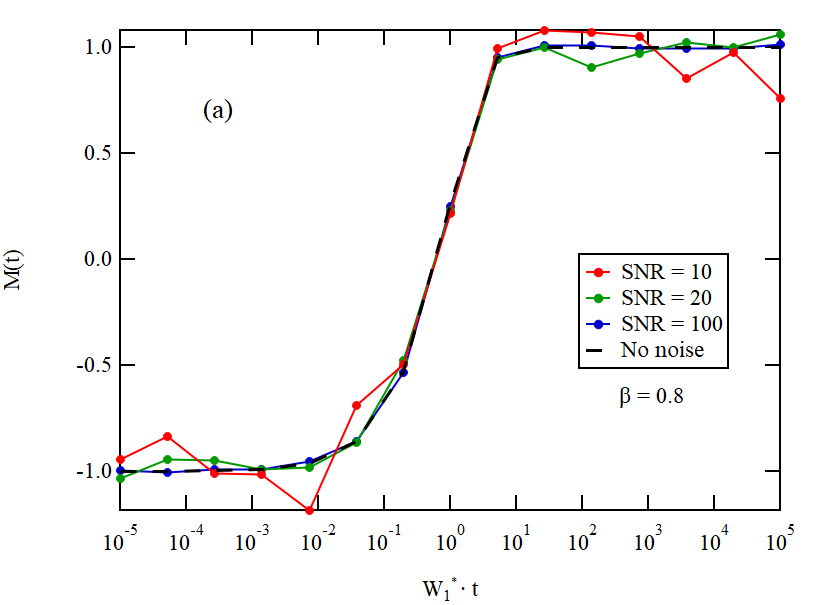}
\includegraphics[width=\linewidth]{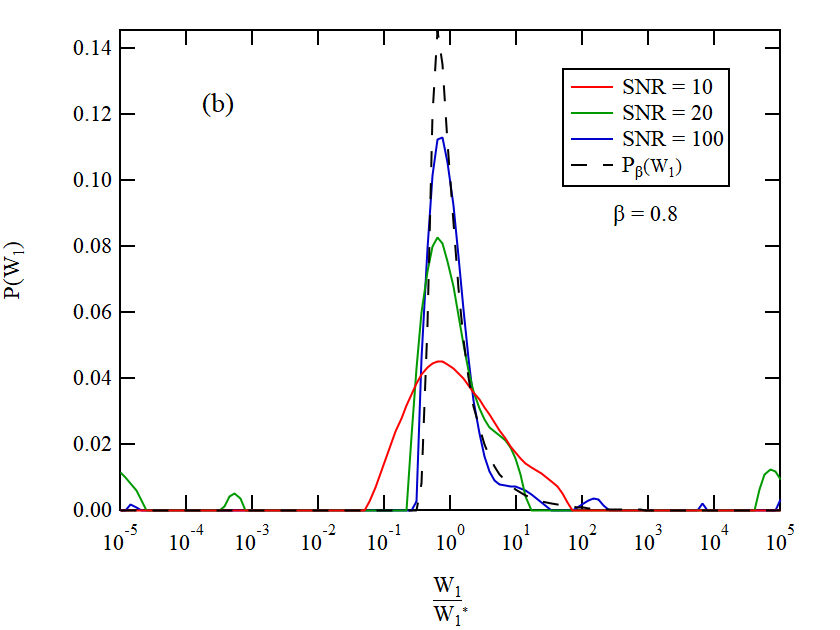}
\caption{(a) Linear-log plot of M(t) given by Eq. \ref{eqn:mtspinhalf} with $N = 15$ time points, SNR = 10, 20, 100, and \textbeta{ } = 0.8. (b) Linear-log plot of comparison between P(W\textsubscript{1}) ILT estimations of M(t) in (a) and the theoretical distribution $P_{\beta}(W_1)$ given by Eq. \ref{eqn:johnstondistribution} for \textbeta{ } = 0.8.\label{fig:ILTnoisy}}
\end{center}
\end{figure}

\begin{figure}[h]
\begin{center}
\includegraphics[width=\linewidth]{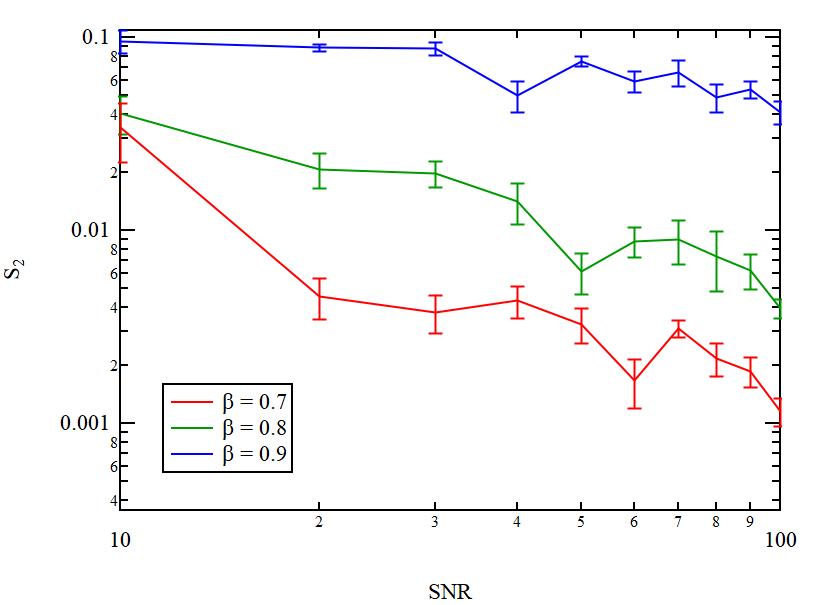}
\caption{\label{fig:S2vsSNR} Log-log plot of $S_2$, given by Eq. \ref{S_2} of P(W\textsubscript{1}) ILT estimations of M(t) given by Eq. \ref{eqn:mtspinhalf} with $N = 15$ time points and SNR ranging from 10 to 100 versus SNR for \textbeta{ } = 0.7, 0.8, 0.9.}
\end{center}
\end{figure}

In order to understand the effect of experimental noise, we added random values $e_i$ to each $M_i$ value, where each $e_i$ is sampled from a Gaussian distribution centered at zero with second moment $\sigma_n^2$, such that the signal to noise ratio $SNR = \sigma_n^{-1}$. A sample set of magnetization recovery points $(N=15)$ with $\beta = 0.8$, and the corresponding ILT distributions are shown in Fig. \ref{fig:ILTnoisy}.
Introducing noise clearly affects the ability of the ILT algorithm to accurately reproduce the known distribution. As shown in Fig. \ref{fig:S2vsSNR}, there is an approximate power law relationship:  $S_2 \sim SNR^{-s}$, where $s$ increases as $\beta$ decreases. Extra structures, such as spurious peaks and shoulders are apparent in the ILT distributions.  These artifacts persist even up to SNR levels of 100, although the overall shape of the distribution is qualitatively correct.  

\subsection{Relaxation of Higher Spins}
For spins greater than $I=1/2$, the kernel function consists of multiple exponential decays reflecting the normal modes relaxation of the spin system. If the nuclei experience only a Zeeman interaction, the energy splittings between the states are all equal. However, it is common that a quadrupolar interaction will further split these states such that the spectrum consists of $2I$ resonances \cite{CPSbook}. For the central ($I_z=+1/2 \leftrightarrow -1/2$) transition in the presence of magnetic fluctuations,  $K(W_1,t) =  1-\phi f(W_1 t)$, where $f(x)$ is given by:
\begin{equation}
   f(x) = \sum_j c_j \exp({-\alpha_j x})
   \label{eqn:highspinkernel}
\end{equation}
and the coefficients $c_j$ are given in Table \ref{table}, and the $\alpha_i$ are $\alpha_1 = 1$, $\alpha_2 = 6$, $\alpha_3 = 15$, $\alpha_4=28$ and $\alpha_5 = 45$.

\begin{table}%[tbhp]
\centering
\caption{\label{table} Coefficients in $f(x)$ for higher spins.}
%\begin{ruledtabular}
\begin{tabular}{|l|c|c|c|c|}
\hline
 & $I={3}/{2}$ & $I={5}/{2}$ & $I={7}/{2}$ & $I={9}/{2}$ \\
\hline
\hline
$c_1$ & ${1}/{10}$ & ${1}/{35}$ & 1/84 & 1/165 \\
\hline
$c_2$ & 9/10 & 8/45 & 3/44 & 24/715 \\
\hline
$c_3$ & 0 & 50/63 & 75/364 & 6/65 \\
\hline
$c_4$ & 0 & 0 & 1224/1716 & 1568/7293\\
\hline
$c_5$ & 0 & 0 & 0 & 7938/12155\\
\hline
\end{tabular}
%\end{ruledtabular}
\end{table}

\begin{figure}[h]
\begin{center}
\includegraphics[width=\linewidth]{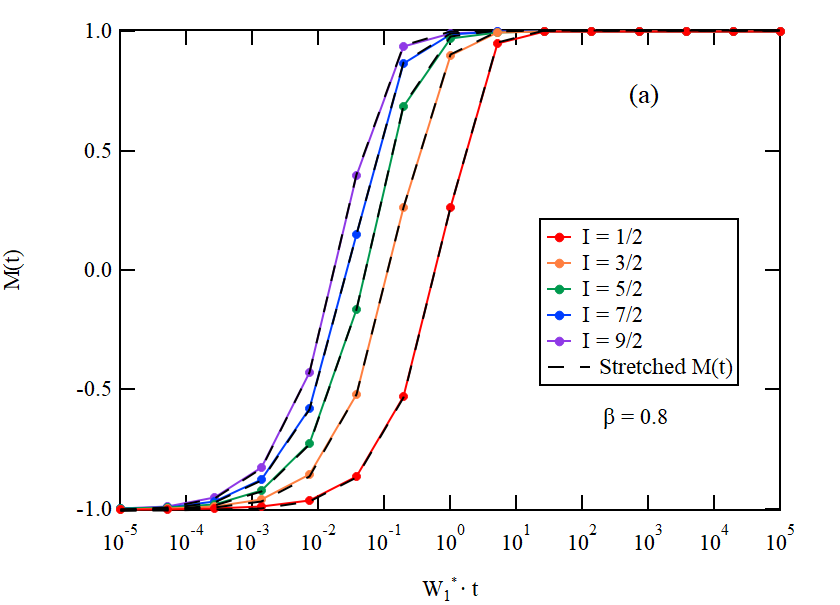}
\includegraphics[width=\linewidth]{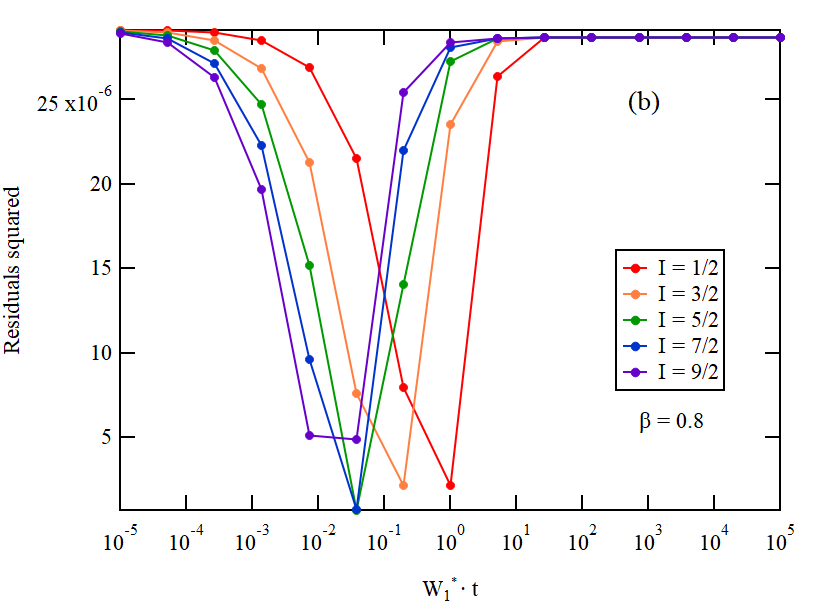}
\caption{(a) Comparison between the computed magnetization recovery $M(t)$ (Eq. \ref{eqn:magrec}) with $P_{\beta}(W_1)$ given by Eq. \ref{eqn:johnstondistribution} for various values of the nuclear spin $I$ with N = 15 time points. (b) Corresponding residuals squared between $M(t)$  and the stretched exponential recovery function (Eq. \ref{eqn:highspinstretch}).
\label{fig:stretchcomparison}}
\end{center}
\end{figure}

\begin{figure}[h]
\begin{center}
\includegraphics[width=\linewidth]{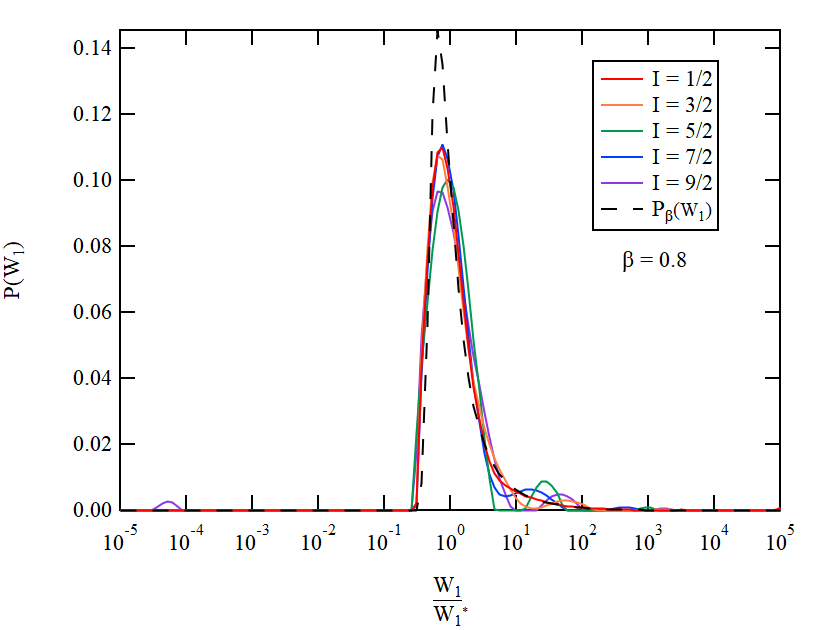}
\caption{Distributions of $P(W_1)$ extracted from stretched relaxation curves using $f_{\beta}(x)$ (Eq. \ref{eqn:highspinstretch}) for $I=1/2, 3/2, 5/2, 7/2$ and 9/2, for $\beta=0.8$.  The dashed line is the theoretical distribution $P_{\beta}(W_1)$(Eq. \ref{eqn:johnstondistribution}). \label{fig:highspindistributions}}
\end{center}
\end{figure}

For stretched relaxation, however, it is unclear how these functions should be modified.  Past researchers have tended to either modify each exponential term with the same stretching exponent:
\begin{equation}
  f_{\beta}(x) = \sum_j c_j \exp({-(\alpha_j x)^{\beta}}),
  \label{eqn:highspinstretch}
\end{equation}
or simply use the spin-1/2 expression of Eq. \ref{eqn:spinhalf} \cite{Curro2000b,MitrovicGlassy214PRB2008,Ba122ClusterGlassNMR}. The problem with these \emph{ad hoc} approaches is that $P(W_1)$ should be independent of the nuclear spin so that it reflects the intrinsic dynamics of the environment, but it is unclear what fitting function should be used.  In order to better understand the distributions for higher spin nuclei, we convoluted  $P_{\beta}(W_1)$ in Eq. \ref{eqn:johnstondistribution} for spin 1/2 with the various Kernel functions in Eq. \ref{eqn:highspinkernel} for different $I$, and compared with Eq. \ref{eqn:highspinstretch}, as shown in Fig. \ref{fig:stretchcomparison}.  Surprisingly, there is near perfect agreement between the two curves, indicated by the low values of residuals squared across all time points. The larger values of residuals squared at the beginning and end of the time domain are most likely due to the limits of the computed magnetization recovery using Eq. \ref{eqn:mtspinhalf} as the exponent term prohibits the recovery to completely reach values of -1 and 1. We further analyzed the various decay curves given by $f_{\beta}(x)$ (Eq. \ref{eqn:highspinstretch}) with the ILT algorithm using the appropriate kernels (given by Eq. \ref{eqn:highspinkernel}) to extract $P(W_1)$ distributions for each value of $I$, as shown in Fig. \ref{fig:highspindistributions}.  Although there are oscillations introduced by the ILT algorithm, the general shape of the distributions for all of the spins are similar to one another and well-described by $P_{\beta}(W_1)$.

\begin{table}%[tbhp]
\centering
\caption{\label{table2} Coefficients in $f(x)$ for different satellite transitions of a spin 7/2 nucleus for magnetic fluctuations.}
%\begin{ruledtabular}
\begin{tabular}{|l|c|c|c|}
\hline
 & $|{1}/{2}|\leftrightarrow|3/2|$ & $|{3}/{2}|\leftrightarrow|5/2|$ & $|{5}/{2}|\leftrightarrow|7/2|$   \\
\hline
\hline
$c_1$ & 196/429 & 49/429 & 4/429  \\
\hline
$c_2$ & 49/132 & 49/132 & 3/44 \\
\hline
$c_3$ & 1/1092 & 100/273 & 75/364  \\
\hline
$c_4$ & 9/77 & 25/308 & 25/77 \\
\hline
$c_5$ & 1/33 & 1/132 & 3/11 \\
\hline
$c_6$ & 1/84 & 1/21 & 3/28 \\
\hline
$c_7$ & 1/84 & 1/84 & 1/84 \\
\hline
\end{tabular}
%\end{ruledtabular}
\end{table}

\begin{figure}[h]
\begin{center}
\includegraphics[width=\linewidth]{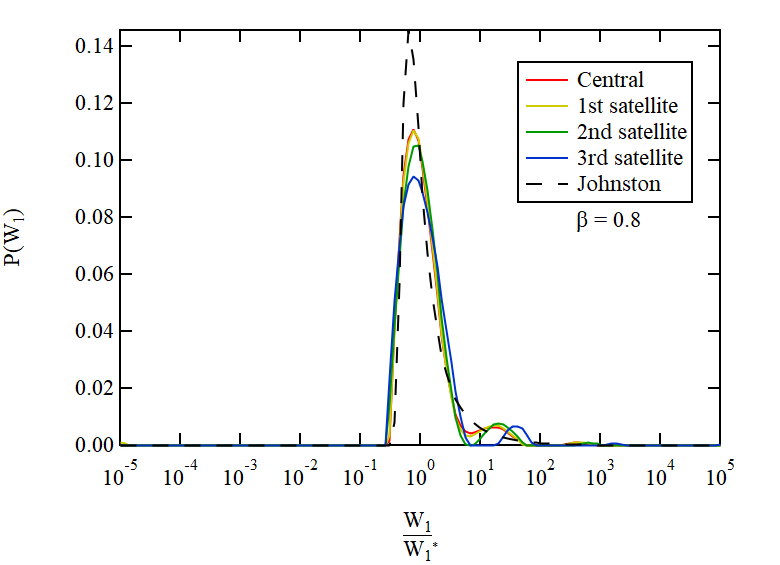}
\includegraphics[width=\linewidth]{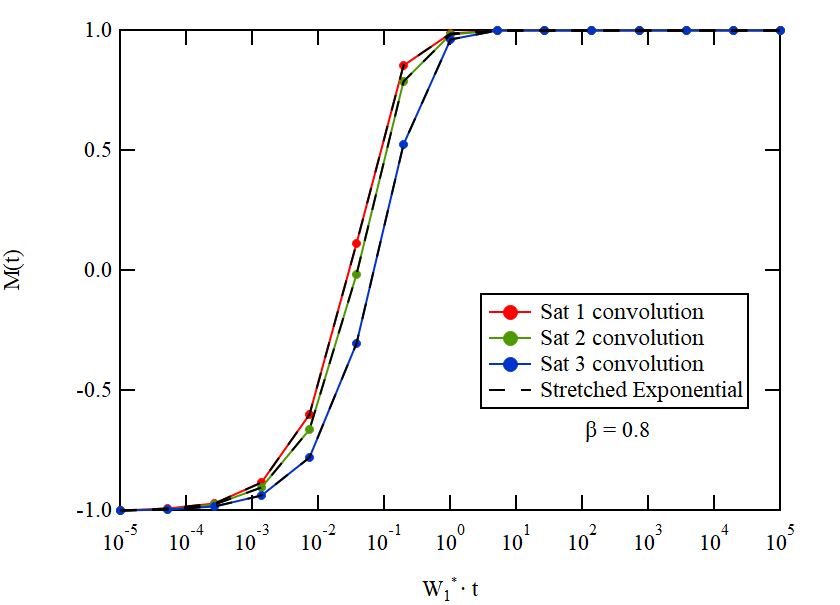}
\caption{ (Upper panel) Distributions extracted from stretched relaxation curves using $f_{\beta}(x)$ (Eq. \ref{eqn:highspinstretch}) for the central ($1/2\leftrightarrow -1/2$), first ($|1/2|\leftrightarrow|3/2|$), second ($|3/2|\leftrightarrow|5/2|$) and third satellites ($|5/2|\leftrightarrow|7/2|$) for $I=7/2$. (Lower panel) Comparison between the computed magnetization recovery and the stretched expression (Eq. \ref{eqn:highspinstretch}) for the same transitions for $\beta = 0.8$.
\label{fig:satellites}}
\end{center}
\end{figure}

Although the relaxation function in Eq. \ref{eqn:highspinkernel} for higher spins is multiexponential, for the central transition the coefficients $c_j$ are such that the relaxation is dominated by one exponential.  On the other hand, for the satellite transitions ($|I_z|  \leftrightarrow |I_z| - 1$, with $1/2 < |I_z| \leq I$), the relative weights of the different exponentials are more evenly distribution.  The coefficients $c_j$ are given for the different satellite transitions in Table \ref{table2} for the case of $I=7/2$.  In this case the $\alpha_i$ are given by $\alpha_1 =1$, $\alpha_2 = 3$, $\alpha_3 = 6$, $\alpha_4=10$, $\alpha_5 = 15$, $\alpha_6 = 21$, and $\alpha_7 = 28$.  Figure \ref{fig:satellites} compares the extracted distributions with $P_{\beta}(W_1)$, and the magnetization recovery using the convoluted $P_{\beta}(W_1)$ with the stretched expression, Eq. \ref{eqn:highspinstretch}, for the central and three satellite transitions for $I=7/2$.  Once again, there is good agreement. These studies indicate that Eq. \ref{eqn:highspinstretch} is the proper form for stretched exponential relaxation so that the distribution is independent of nuclear spin, $I$.

\section{Discussion}
The ILT algorithm appears to be most effective at accurately capturing the true distribution of relaxation rates when the distribution is sufficiently broad to begin with. For stretched exponential relaxation, when $\beta \geq 0.8$, or when the width of the distribution is less than about one decade, the ILT algorithm overestimates the width.  This observation reflects that fact that the Tikhonov regularization acts to smooth the distribution. Efforts to invert noisy data of ill-posed problems typically result in large fluctuations of the distribution function that are not physical, hence the effort to `regularize' the solution \cite{butler1981estimating,lawson1995solving}.  Smoothing of a distribution is a necessary side-effect of the ILT algorithm, and will lead to overestimates of the distribution width when the distribution is intrinsically narrow (such that $\beta \geq 0.8$).

Choosing the optimal number of measured recovery points, $N$, is important to accurately capture the distribution, and our simulations indicate that $N$ should be at least 12-15.  Choosing a greater number of points improves the accuracy, but may lead to a reduction in signal to noise if the total experimental time is constrained. Signal to noise ratios above $\sim 40$ are necessary to capture the salient features of a distribution, but we find that unphysical artifacts in the distribution persist even up to higher SNR values. This fact should be taken into account when interpreting distributions obtained from experimental systems.

The ILT algorithm artificially introduces discontinuities in $dP(W_1)/dW_1$ when $P(W_1)$ approaches zero because the Heaviside function forces $\vec{P}$ to vanish if it becomes negative. This behavior is somewhat problematic because it would be more physically-realistic if the tails of $P(W_1)$ asymptotically approached zero smoothly. Such artificial cut-offs may not accurately capture the physics of glassy systems where the fluctuations are expected to exhibit power law distributions with long tails \cite{sollich1997rheology,fujisaka1993glassy}.   However, our analysis clearly demonstrates that fitting the magnetization recovery directly with the stretched exponential expression (Eq. \ref{eqn:highspinstretch}) and inferring the distribution using the theoretical expression (Eq. \ref{eqn:johnstondistribution}) provides a straightforward description of the distribution of relaxation rates.  This approach is valid for any nuclear spin, is easier to implement, and does not suffer from the introduction of artifacts.  In such cases, all of the relevant physical information about the distribution is captured by the parameters $W_1^*$ and $\beta$. Note that the ILT approach may still be necessary for cases in which the distribution is not well-described by a stretched exponential, for example a bimodal distribution of relaxation rates, or when the distribution is not expected to be well-approximated by $P_{\beta}(W_1)$.  The latter distribution is biased towards high $W_1$ values, and thus stretched exponential fits would not be appropriate when the distribution is expected to be more symmetric or biased towards low $W_1$ values.

\section{Conclusion}

The ILT algorithm is a powerful method to extract distributions from  time-series data sets, which has grown in popularity in recent years.  A priori, this method makes no assumptions about the nature of the distribution, and is thus useful to study materials with complex inhomogeneous behavior.  However, the algorithm does 'filter out' sharp features of a distribution, leading to artificial broadening and oscillatory components.  These features are especially pronounced when the time-series data has significant levels of noise.  On the other hand, many researchers have traditionally fit the time-series data directly with stretched exponentials of various forms, which are easier to implement and more direct.  A drawback of this method has been poor understanding of the nature of the distribution, particularly for the case of $I>1/2$.  Our study indicates that the stretched exponential form described by Eq. \ref{eqn:highspinstretch} accurately captures the distribution \emph{independent} of the spin $I$.  This result implies that if the time-series data of any nucleus can be fit by this form, the full distribution $P(W_1)$ can be inferred without the need to invert the data with a complicated algorithm such as the ILT.

\textit{Acknowledgment.} We acknowledge stimulating discussions with W. Polonik, K. Dahmen and E. Carlson.  Work at UC Davis was supported by the NSF under Grants No. DMR-1807889 and PHY-1852581.

\bibliography{ILTbibliography}

\end{document}